\documentclass[%
superscriptaddress,
amsmath,
amssymb,
]{revtex4-2}

\usepackage{caption}
\usepackage{subcaption}
\usepackage{graphicx}% Include figure files
\usepackage{dcolumn}% Align table columns on decimal point
\usepackage{bm}% bold math
\usepackage{color, xcolor}
\usepackage{booktabs}
\usepackage{amsmath}
\usepackage{siunitx}
\usepackage{comment}
\usepackage[T1]{fontenc}
\usepackage{minted}
\usepackage{tikz}
\usepackage{booktabs} % nice rules, optional
\usepackage{tabularx}
\usetikzlibrary{shapes.multipart, positioning}
\usepackage{makecell}
\usepackage[colorinlistoftodos]{todonotes} % adds support for todo notes
\usepackage{lipsum}
\usepackage{hyperref}
\hypersetup{
  colorlinks=true
}

\begin{document}

%\preprint{APS/123-QED}

\title{Closing the Loop: Deploying Auto‑Generating Digital Twins for Particle Accelerators}

\author{A.~D.~Brynes}\email[]{alexander.brynes@stfc.ac.uk}
\affiliation{Accelerator Science \& Technology Centre, STFC Daresbury Laboratory, Warrington, United Kingdom}
\affiliation{Cockcroft Institute, Warrington, United Kingdom}
\author{M.~King}\email[]{matthew.king@stfc.ac.uk}
\affiliation{Accelerator Science \& Technology Centre, STFC Daresbury Laboratory, Warrington, United Kingdom}
\affiliation{Cockcroft Institute, Warrington, United Kingdom}
\author{K.~R.~L.~Baker}
\affiliation{ISIS Neutron and Muon Source, STFC Rutherford Appleton Laboratory, Harwell Campus, Didcot, United Kingdom}
\author{R. Banerjee}
\affiliation{ISIS Neutron and Muon Source, STFC Rutherford Appleton Laboratory, Harwell Campus, Didcot, United Kingdom}
\author{R.~Clarke}
\affiliation{Technology Department, STFC Daresbury Laboratory, Warrington, United Kingdom}
\author{D.~J.~Dunning}
\affiliation{Accelerator Science \& Technology Centre, STFC Daresbury Laboratory, Warrington, United Kingdom}
\affiliation{Cockcroft Institute, Warrington, United Kingdom}
\author{J.~K.~Jones}
\affiliation{Accelerator Science \& Technology Centre, STFC Daresbury Laboratory, Warrington, United Kingdom}
\affiliation{Cockcroft Institute, Warrington, United Kingdom}
\author{M.~Leputa}
\affiliation{ISIS Neutron and Muon Source, STFC Rutherford Appleton Laboratory, Harwell Campus, Didcot, United Kingdom}
\author{A.~E.~Pollard}
\affiliation{Accelerator Science \& Technology Centre, STFC Daresbury Laboratory, Warrington, United Kingdom}
\affiliation{Cockcroft Institute, Warrington, United Kingdom}
\author{M.~Romanovschi}
\affiliation{ISIS Neutron and Muon Source, STFC Rutherford Appleton Laboratory, Harwell Campus, Didcot, United Kingdom}
\author{M.~Shaw}
\affiliation{Technology Department, STFC Daresbury Laboratory, Warrington, United Kingdom}
\author{N.~Ziyan}
\affiliation{Accelerator Science \& Technology Centre, STFC Daresbury Laboratory, Warrington, United Kingdom}
\affiliation{Cockcroft Institute, Warrington, United Kingdom}

\date{\today}

\begin{abstract}
The simulation of a physical system in a virtual replica, known as a digital twin, is a useful way to interrogate the system non-invasively, providing the ability to perform predictive maintenance and surveillance, and to investigate potential novel configurations without perturbing the system. This article presents the implementation of an auto-generating digital twin architecture for particle accelerators: a virtual control system is generated to mirror the physical accelerator hardware, and used to update a simulation model which then feeds back the results into virtual diagnostics. All of the information about the accelerator lattice is cascaded down from a ground source of truth, removing any ambiguity about the naming of parameters between the simulation model and the virtual hardware. This design is modular and extensible, allowing researchers from different institutions to use their own models (for example, a machine learning model) and accelerator lattices while maintaining the overall structural coherence of the digital twin. This architecture has been tested for three accelerator facilities \textendash~CLARA, the ISIS injector, and the proposed UK XFEL \textendash~and aims to provide the foundation for a collaborative community effort in the development of shared technology towards a generic digital twin solution. 
\end{abstract}

\maketitle

\iffalse
\textbf{Top-level overview of what should be in the paper}
\begin{itemize}
    \item Introduction to digital twins
    \begin{itemize}
        \item Why they are useful in general
        \item How they apply to accelerators
    \end{itemize}
    \item Components
    \begin{itemize}
        \item Generic lattice format (LAURA) including control system description and simulation export format
        \item Simulation framework (SIMBA)
        \item Virtual accelerator construction (SARABI)
        \item Control system interface (CATAP)
    \end{itemize}
    \item Architecture
    \begin{itemize}
        \item Controls layer
        \item Simulation layer
        \item Common interface
        \item Lattice API
        \item Job scheduling / Kafka
    \end{itemize}
    \item Deployment
    \begin{itemize}
        \item CLARA / ISIS
        \item Interface / GUI
    \end{itemize}
\end{itemize}
\fi

\section{Introduction}

A digital twin (DT) refers to an integrated simulation of a complex process using the best available models to mirror its physical state \cite{DigitalTwin,RemoteSens.14.6.1335,ApplSystInnov.4.2.36,ComputInd.130.103469}. Since its original formulation, dozens of definitions of a DT have been proposed \cite{ComputInd.130.103469}, yet they all share common features: the status of a physical mechanism is recorded; this data is fed into a high-fidelity model of the process; this simulation is used to provide information about the process and to help inform the course of action. The widespread interest in DTs across such disparate fields as manufacturing \cite{JCompDesEng.1.3.213}, construction \cite{AutomConstr.114.103179}, healthcare \cite{FrontGenet.9.31}, and product design \cite{IntJProdRes.57.12.3935} highlights their generic utility. 

Physics is a field that is well-suited for DT integration. The operation and characterization of complex physical systems often relies on measurements from a wide range of instruments, and many aspects of the system are typically difficult to measure, particularly in a non-invasive way. As such, computer simulations can be a great benefit to researchers aiming to understand or optimize the system. DTs have been deployed to monitor and optimize physical systems such as nuclear reactors \cite{NuclSciEng.196.6.668} and wind farms \cite{EnergyConversManag.293.117507}. One potential issue with using standard simulation tools as the basis for a DT is the computational intensity of the task: depending on the complexity of the physical system, modelling can require a great deal of resource in terms of time or compute. Physics-informed neural networks can therefore provide a realistic path towards the deployment of a real-time DT across a wide domain \cite{arXiv.2401.08667}.

A particle accelerator is an ideal system for the development of a physics-informed DT \cite{arXiv.2507.20493}. These machines are composed of a wide variety of components used for the control and monitoring of particle beams. Physicists understand and model these beams as collections of particles in a six-dimensional phase space which has three dimensions each of position and momentum; components in the accelerator then control the phase spaces of these beams. In practice, accelerators are typically operated via a control system, which provides an interface to the physical hardware. Measurements of the beam can be either invasive or non-invasive, and in most cases they provide a reduced representation of the full 6D phase space of the particle beam. There is therefore a potentially significant gap between measurements in the control system and the physicist's model of the accelerator. Simulation codes and machine learning (ML) models are widely used for bridging this gap, with control system parameters being converted into units that are suitable for modeling the physical setup of the accelerator and comparing with measurements. Oftentimes this comparison is done by hand, and the data (measured or simulated) can require adjusting in order to provide a fair comparison. 

A DT for an accelerator should aim to reproduce, as closely as possible, the configuration of the machine in real time, making use of either conventional simulation codes or an ML-based model. A virtual replica of the control system should be updated based on the physical system and used to track particle beams, and then updated based on the outputs of the model. In this way, a fully integrated DT can be used to provide physicists and operators with real-time (or near to real-time) estimates of the physical state of the machine, opening the door to predictive non-invasive measurements, a framework for the development of offline optimization strategies, and a virtual commissioning suite. 

Strictly speaking, a DT can only be considered as such if there is a bi-directional data linkage between the digital and physical representations of the system \cite{BusInfSystEng.64.375,AdvModelSimulEngSci.7.13}; in other words, the results produced by the DT are used directly (and autonomously) to update the physical system. A tool that has only a uni-directional data flow from a physical to virtual representation -- with the potential option to update the physical system manually -- is better described as a digital shadow (DS). The DS can be understood as a propaedeutic stage in the development of a DT, providing the observational and data-infrastructural foundations required for bi-directional coupling and intervention, while first retaining the ability to check the validity of the model results, and providing the opportunity for important decisions to be made by operators.  

This article outlines the components, architecture and initial deployment of a particle accelerator DT. Its foundation is based on a comprehensive description of the elements of the accelerator lattice, incorporating control system information, physical attributes, and simulation code-specific parameters (Sec.\,\ref{sec:architecture}). This is used to construct a virtual copy of the control system, which can then be modified based on a particular setup of the accelerator. A configurable simulation toolkit is then used to execute the simulation and to return the predicted beam evolution, and full output distributions at specified locations, back into the simulated control system. 
%A diagram representing the flow of information for this DS is shown in Fig.\,\ref{fig:flowchart}. 
Each component of this framework is modular, with the virtual control system, simulation model, and communication layer placed in separate Docker containers. This facilitates the substitution of each module with a different container, depending on the particular use-case. Our framework has been deployed on the CLARA accelerator \cite{PhysRevAccelBeams.23.044801} and tested on the ISIS Neutron and Muon Source \cite{NIMA.917.61} Virtual Injector, and used for developing a virtual commissioning procedure on the proposed UK XFEL facility \cite{marangos20, UKXFEL} (Sec.\,\ref{sec:deployment}). Given the generalizable nature of each aspect of this framework, this architecture provides a suitable path for the full exploitation of DT technology by the accelerator community. Some of the possible routes this development could take are discussed in Sec.\,\ref{sec:conclusions}.

\iffalse
\begin{figure}
    \centering
    \includegraphics[width=0.6\linewidth]{figures/integrated-model-flowchart.png}
    \caption{Diagram representing the flow of information and functionality of the DT. The loop begins with the reading of the lattice settings -- automatically or in `triggered' mode -- which update the \texttt{Lattice} object and send it to the communications layer. This is then queried by the simulations layer, which checks if those settings have previously been tracked. If they have, then no action is taken by the simulation module; if the settings are new, they are tracked, and the outputs from the simulation are also written to the \texttt{Lattice} object and the virtual control system. Finally, this updated object is saved to the database in the communications layer, and the loop starts over.}
    \label{fig:flowchart}
\end{figure}
\fi
\section{Architecture}\label{sec:architecture}

\begin{figure}
    \centering
    \includegraphics[width=\linewidth]{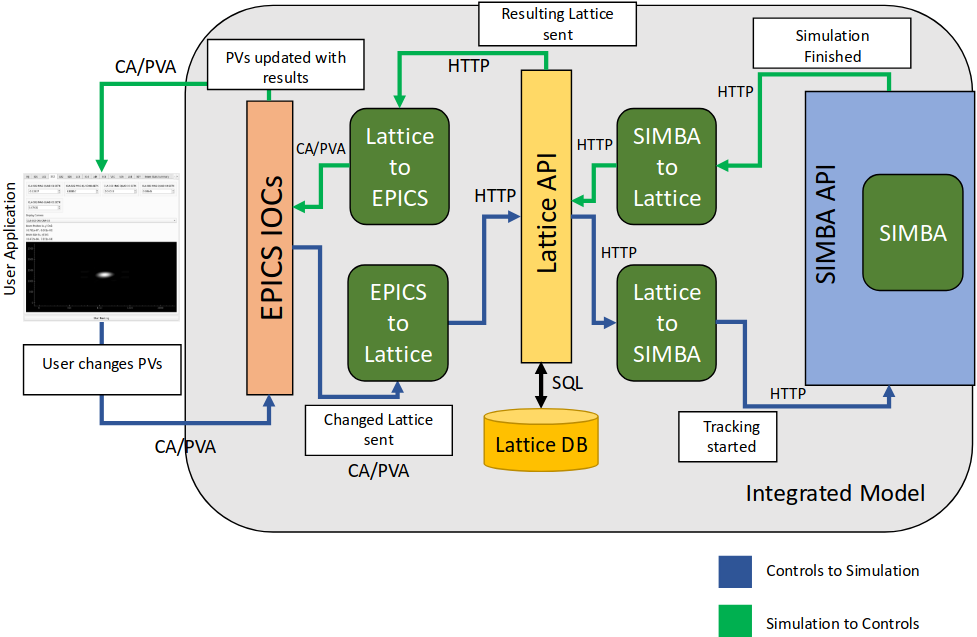}
    \caption{Architecture of the DT. The communications layer facilitates the transfer of information about a full accelerator lattice between a virtual control system and a simulation module, separating out the elements of the DT while maintaining a coherent and consistent flow of information between layers. Communication between modules is either done via HTTP, or using EPICS Channel Access (CA) or PV Access (PVA) for control system interactions.}
    \label{fig:architecture}
\end{figure}

Structurally, the DT consists of a collection of Docker containers that execute their own internal functions and communicate with each other via a central controller. This modular architecture is useful because it both separates out the functionality, and allows modules to be swapped out depending on the use-case. Each module can be accessed via \texttt{FastAPI} \cite{FastAPI}, with API calls linked to internal functions of that module. The separate Docker containers expose these API calls to other modules, allowing a strict separation of functionality while maintaining the overall coherence of the system. The architecture of the DT is outlined in Fig.\,\ref{fig:architecture}, while the functionality of the various modules and elements are described in detail in the sections below.

\subsection{Lattice}\label{subsec:lattice}

A coherent description of a complex system such as a particle accelerator is dependent on a ground source of truth about the system. Simulation models represent information about accelerator elements in a different way to control system variables, and a format that can incorporate both of these descriptions is fundamentally important to the development of a DT. The \texttt{LAURA} format \cite{LAURA} (Lattice Architecture for a Unified Representation of Accelerators) provides this comprehensive description: each accelerator element is instantiated from a file that contains all relevant information required to construct both a representation of that element in a control system, and the physical attributes of that element. For an accelerator magnet, the magnetic strength, length, and position are defined, along with the control system parameters used to adjust the current/strength/field in the magnet, and other information. From this description, the system has access to both sets of information pertaining to all elements in the lattice. 

Each element in the lattice contains an internal representation describing the name and type of the element, along with the ability to include summary information about the beam at that location such as Twiss parameters, centroids and widths in 6D phase space. Specific elements, including magnets and RF cavities, for example, rely on additional properties such as magnetic strength, or phase and field amplitude, respectively. At certain locations in the lattice, for example at diagnostic stations, it is useful to store a full multiparticle representation of the bunch in 6D phase space, so these are also included in the \texttt{Lattice} object. The entire lattice is built from \texttt{Section} objects -- defined via \texttt{LAURA}, each of which contain a universally unique identifier (UUID) for a particular setup. A generic beam summary object, which is a top-level overview of the beam evolution, can also be associated with the lattice. Essentially, the internal representation of the lattice passed between modules is a reduced version of the full \texttt{LAURA} lattice with beam data added, containing only information that is updated by each of the modules; sending the full lattice provided by the \texttt{LAURA} instance could introduce an unnecessary overhead in terms of both computation and complexity. 

\subsection{Communications Module}\label{subsec:comms}

The DT modules are then used to update the \texttt{Lattice} object: depending on their function, each module is used either to update the input lattice, or to produce an output. This is handled by a communications layer that facilitates the passing of the \texttt{Lattice} object between the simulation and controls modules. At the end of the loop, the full lattice represents both the state of the machine hardware (the simulation input) and the evolution of the particles (the simulation output). This full \texttt{Lattice} object is then stored in a database which can be accessed for post-processing. 

%\todo[inline]{Update with description of kafka once it's done.}

Two operational modes are provided, and can be toggled in the virtual control system: continuous and triggered. In continuous mode, the communications layer periodically checks if any of the specified control system variables have been changed (see Sec.\,\ref{subsec:catap} below); if they have, the \texttt{Lattice} object is updated and passed to the simulation module (see \ref{subsec:simulation} below). Alternatively, in triggered mode, the virtual control system can be updated offline, and only once the user triggers a simulation will it be executed. 

\subsection{Virtual Accelerator}\label{subsec:virtual_accelerator}

Given that control system information is also provided in the \texttt{LAURA} schema, an instance of the lattice can be used to construct a virtual replica of the physical control system, for all of the control system variables that are defined. \texttt{SARABI} (Soft Architecture for Rendering Automated Backend IOCs) is a Python package designed to create virtual soft Input/Output Controllers (IOCs) for the EPICS (Experimental Physics and Industrial Control System) control system \cite{EPICS} from YAML configuration files. It provides a flexible architecture for creating IOCs programmatically, making it easier to manage and simulate EPICS records. A schema translator is used to map YAML data to EPICS records; the package is extendable to support various device types and configurations, and EPICS environment variables are configured for seamless integration. 

Using the full lattice definition files containing control system information, hardware types are grouped together, and their associated controls variables are extracted. The control system identifier (the EPICS Process Variable, or PV, in our case), data type, controls protocol, description, and middle layer handle for each variable are fed into a \texttt{Jinja2}~\cite{Jinja2} template file to construct the code for a virtual IOC. A base-level IOC is produced for each hardware type, and the channels for each element are constructed for each element of that type. All of the IOCs can then be launched to generate a full replica of the control system described in the lattice files. In order to separate out the virtual and physical control systems, and to avoid naming conflicts, every control system variable defined in the \texttt{LAURA} lattice is prepended with the prefix \texttt{VM-} in the virtual IOCs. It should be noted that while this virtual accelerator is based on EPICS, the \texttt{LAURA} framework allows for the definition of element control variables in any other control system such as TANGO \cite{TANGO}; an equivalent module to \texttt{SARABI} could easily be developed to generate virtual TANGO servers, and an appropriate TANGO-based middle layer package would be required for the control system interface (Sec.\,\ref{subsec:catap}).

Currently, the virtual IOC functions only as a basic I/O control system, meaning that PVs are not linked together to trigger distributed updates across the control system. On a real control system, for example, updating the set current on a magnet may also trigger an update to the magnet strength parameter, based on calibration factors. These expressions can be defined via the \texttt{LAURA} schema, and can in principle be used to construct a more detailed and accurate representation of the real control system. Alternatively, a more realistic virtual control system can be prepared offline and imported into the DT on instantiation. 

While the hardware attributes can be constructed based on the variables defined in the \texttt{LAURA} elements, the output variables -- in other words, the simulation results -- can also be created for the virtual accelerator. Every element can be associated with parameters describing the summary information about the beam, such as Twiss parameters and emittance; this provides a control system-centric approach which is useful for comparing physical measurements to simulation outputs, and which is an important feature in the realization of a DT. Specific physics IOCs can be instantiated which associate each element with a number of simulation PVs, prepended with the string \texttt{SIM-}. This is done using the \texttt{p4p} \cite{p4p} package, which allows the creation of controls variables with a range of formats, including 6D phase space arrays associated with diagnostic screens or other critical locations in the beamline. These IOCs are also generated procedurally, based on the lattice elements provided for that instance of the DT.

\subsection{Control System Interface}\label{subsec:catap}

The control system information provided by \texttt{LAURA} elements is also used to inform the other modules about the current state of the accelerator. A streamlined, user-friendly interface to the control system is provided by the \texttt{CATAP} \cite{arXiv.2509.19794} middle layer package. This software library provides methods for generating a full snapshot of the current state of the accelerator (physical or virtual), and therefore can be used to package the information received from the control system into a suitable format that can be passed between modules -- for example, to convert the current in a magnet to its corresponding magnetic strength, or to determine the RF accelerating gradient from the power in the cavity. A similar module to \texttt{SARABI} is available for procedurally generating a middle layer based on definitions in the lattice file \cite{arXiv.2509.19794}; this was done for testing the DT on the ISIS virtual injector, while a more developed version of \texttt{CATAP} was used for the deployment of the DT on CLARA (Sec.\,\ref{sec:deployment}). 

Certain elements, or elements of a given type, can be provided as the input parameters used to update the \texttt{Lattice} object (see Sec.\,\ref{subsec:lattice}); for example, if the strength of a magnet, or the phase of an RF cavity, is changed on the physical accelerator, this information can be sent to the DT to prepare a new simulation run. For initial testing purposes, the following accelerator and beam systems were available as triggers used to update the simulation: quadrupole and dipole magnet strengths; RF cavity phase and power; initial Twiss parameters for each machine section; the simulation code to be used for a given section; and beam generator parameters (i.e. the initial distribution), including total charge, the number of macroparticles, and the average initial beam distribution properties. Depending on the accelerator and the requirements of the DT, any parameter associated with the machine or the beam can be used as a trigger for updating the simulation. 

\subsection{Simulation Framework}\label{subsec:simulation}

The \texttt{LAURA} package also provides functions for exporting sections of an accelerator lattice into the formats required for a number of simulation codes. A variety of multi-particle beam physics modalities are supported across these codes, including low-energy space-charge-dominated beam dynamics, acceleration in radiofrequency cavities or plasmas, and free-electron lasers. Lattice sections can be defined in a file, and each element is exported sequentially.

This lattice model can then be used to construct, execute and process simulations for each section, for example with the \texttt{SIMBA} package \cite{SIMBA} (Simulations for Integrated Modeling of Beams in Accelerators), which uses an instance of a \texttt{LAURA} lattice to execute start-to-end simulations of an accelerator, with seamless switching between codes for different lattice sections. Simulation outputs -- both beam distributions at specified locations, and general summaries of the beam evolution -- are saved to standardized formats. In addition to the standard multi-particle tracking codes supported, \texttt{SIMBA} also provides an interface to ML models of a lattice section via the \texttt{Poly-Lithic} library \cite{PolyLithic}, which enables the deployment of models with arbitrary inputs and outputs. Provided that the model called describes the accelerator elements for that section, and that it returns its output information in a suitable format, \texttt{SIMBA} considers it the same as any other simulation code. 

Once the \texttt{Lattice} object is updated in the communications layer (see Sec.\,\ref{subsec:comms}) above, the simulation module can be called. For our DT, this is based on \texttt{SIMBA}, but in principle any package that is able to model the propagation of a beam through an accelerator and return the appropriate outputs can be used. As with the control system module (see Sec.\,\ref{subsec:catap}), certain elements in the \texttt{Lattice} are specified to be relevant for updating the simulation model; these are then transferred to the instance of \texttt{SIMBA}.

Once updated internally, the input variables for each section are checked sequentially and compared with existing entries in the database (see Sec.\,\ref{subsec:comms}). If the full \texttt{Lattice} exactly matches a run that has been performed before, then no simulation is executed and the virtual control system is updated with outputs from the database. If any of the \texttt{Section} objects have been changed, then the model is updated accordingly and a simulation is run, starting from the section which has changed. The database stores the UUIDs for all previous tracking runs, and \texttt{SIMBA} is able to load in the beam distribution for the beginning of that section for a previous run, avoiding the need to run a full start-to-end simulation if the lattice has changed at some point further on in the accelerator. Another aspect of the simulation model that can be updated via the virtual control system is the tracking method to be used for each section of the lattice: \texttt{SIMBA} supports both a number of tracking codes and the calling of ML models via \texttt{Poly-Lithic}. These settings are also passed through in the \texttt{Lattice} object and can be set via the virtual control system. 

At the end of a simulation run, the simulation outputs are written to the \texttt{Lattice} via the communications module. Summary information of the beam evolution through each section is saved, as are beam properties and phase spaces at given locations, such as screens and markers. This full \texttt{Lattice} object is saved to the database and can be queried via its UUID. Next, these attributes are all written to the virtual control system via the \texttt{SIM} PVs for each element (see Sec.\,\ref{subsec:virtual_accelerator}). Once the full \texttt{Lattice} object (if new) is saved in the database, the loop is completed.

\subsection{Web Interface}\label{subsec:web}
\begin{figure} 
    \centering
    \includegraphics[width=\linewidth]{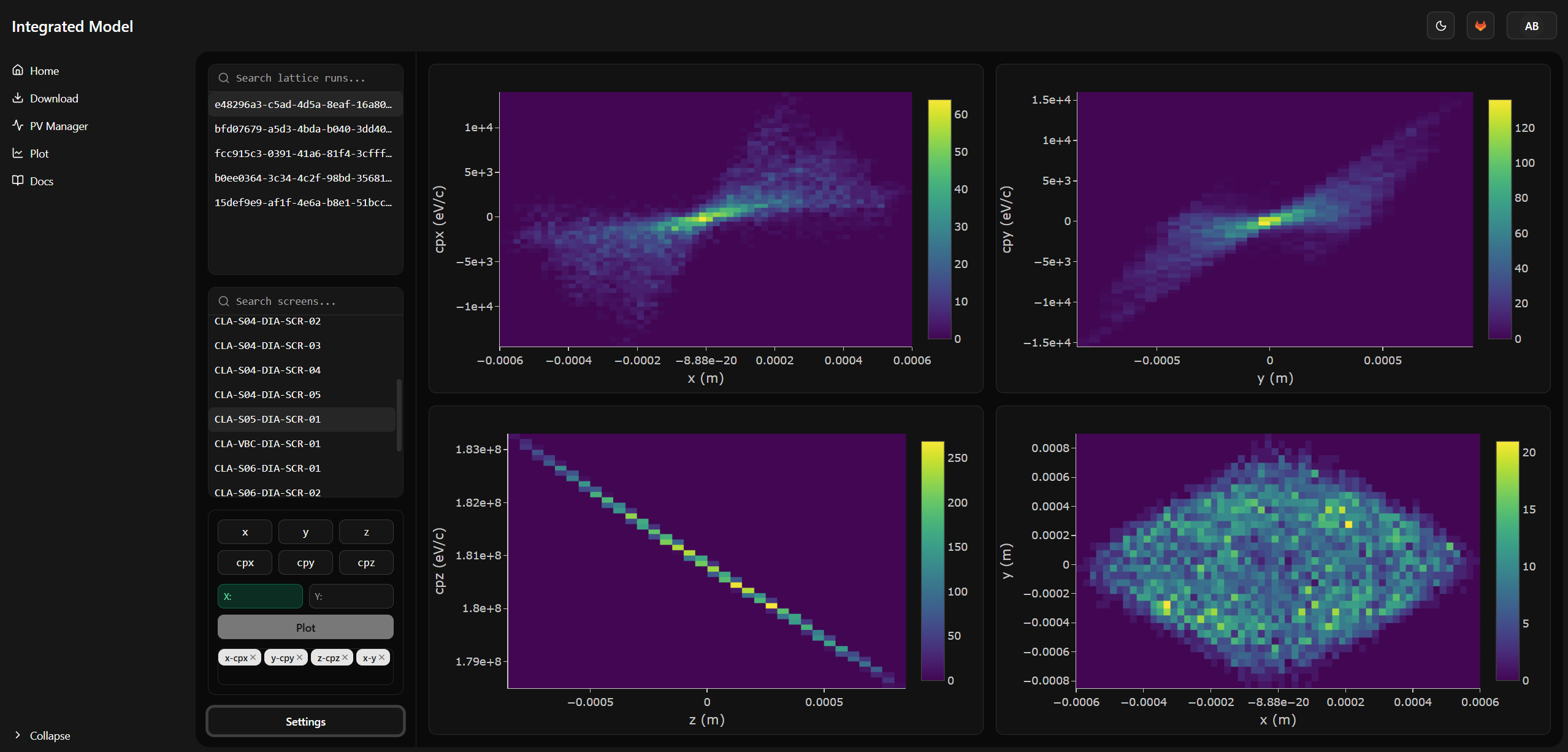}
    \caption{The web interface of the DT, showing the phase-space visualization page. Users select a simulation run, a diagnostic screen location, and pairs of phase-space coordinates to plot. Multiple plots can be displayed simultaneously in a configurable grid layout.}
    \label{fig:web}
\end{figure}

A web-based interface has been developed to provide physicists and operators with a browser-based environment for interacting with the DT. The interface is built using \texttt{React} \cite{React} and \texttt{TypeScript} \cite{TypeScript}, and communicates with the virtual control system via PVWS (Process Variable Web Sockets) \cite{PVWS}, a protocol for exposing EPICS PVs to web clients. This allows users to read, monitor, and write to virtual accelerator PVs directly from the browser, with PV subscriptions handling automatic reconnection and real-time updates. In this way, the full operational loop can be driven from a single interface: users adjust virtual PVs, triggering the simulation pipeline, with results made available for inspection within the same application.
The interface also provides interactive visualization of simulation outputs. Beam distribution data stored in the communications layer database can be retrieved and rendered as phase-space plots. As shown in Fig.\,\ref{fig:web}, users select a simulation run by its UUID, choose a diagnostic screen location along the beamline, and specify the phase-space coordinates to be displayed, either as a two-dimensional density histogram or a scatter plot
%rendered using the Plotly \cite{Plotly} and Apache ECharts \cite{ECharts} libraries, respectively
. Multiple phase-space projections can be displayed simultaneously, providing a convenient way to inspect the full output of a simulation run at any instrumented location in the lattice.

This interface removes the need for command-line interaction with the DT, improving accessibility for users to interrogate simulation outputs and adjust machine settings without requiring knowledge of the underlying infrastructure.

\section{Deployment}\label{sec:deployment}

\subsection{Model-Matching Investigations on CLARA}\label{subsec:clara}

This DT model has been deployed on the CLARA accelerator for initial testing. Machine snapshots are generated using the CATAP middle layer software \cite{arXiv.2509.19794}, which records settings and diagnostic readings for a given machine setup. These snapshots can then be used directly to update the virtual accelerator to trigger a simulation of the experimental setup, and to generate readings on virtual diagnostics. Alternatively, the DT can be run in `live' mode, with updates triggered once a control system parameter is modified. 

\begin{figure}
    \centering
    \includegraphics[width=0.6\linewidth]{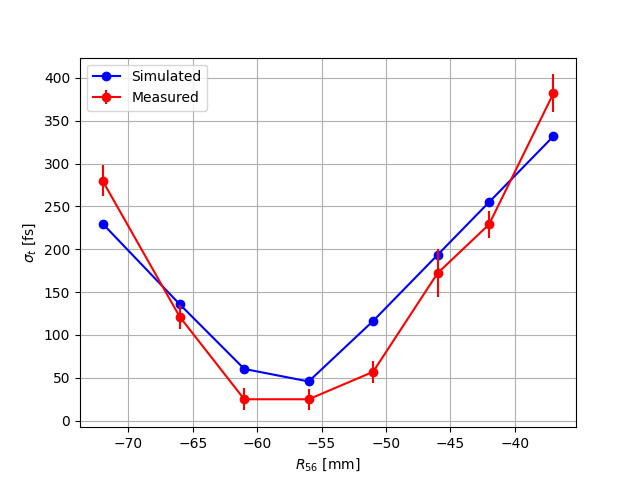}
    \caption{Measurements of the electron bunch length (RMS) in CLARA as a function of the $R_{56}$ of the variable bunch compressor, and simulations of the same setup performed using the DT, based on the machine setup from the virtual control system.}
    \label{fig:bunch-length-comparison}
\end{figure}

An example measurement and its corresponding virtual counterpart are shown in Fig.\,\ref{fig:bunch-length-comparison}: the bunch length of a $3$\,\si{\pico\coulomb} beam was measured using a transverse deflecting cavity at the end of the CLARA linac as a function of the $R_{56}$ of the variable bunch compressor. This measurement was simulated directly in the DT based on physical machine settings and using virtual control system interactions, with the RMS bunch length values written into the virtual control system. Good agreement was found in terms of the trends of the virtual and physical experiments, and the minimum bunch length was found at a similar $R_{56}$. The quantitative agreement is not perfect, which could be attributed to the fact that the beam optics was not optimized at each step of the scan, potentially affecting the reliability of the measurements which rely on knowledge of the transverse beta functions at the deflecting cavity. More detailed studies and machine development time would be needed to reach towards a more optimal agreement. This initial test, however, demonstrates that the DT is readily available to be used for exploratory model-matching experiments, and that it can provide a reasonable estimate of the bunch properties with minimal effort. This can be particularly useful in locations where diagnostics are not available on the physical machine, or for estimating bunch properties that are difficult to measure.

\subsection{Machine Learning Model Testing on ISIS}\label{subsec:isis}

In order to test the deployment of machine learning models in the DT, a simple neural network was trained using simulations from the ASTRA tracking code \cite{ASTRA} to predict the beam evolution through the ISIS Medium Energy Beam Transport (MEBT) section \cite{RevSciInstrum.90.103310}. The model weights and a structured dictionary containing the required inputs (initial beam distributions and settings of the magnets and RF cavities) and expected outputs (the Twiss parameters along the MEBT section) are then sent to the \texttt{Poly-Lithic} server internal to the DT on instantiation. 

\begin{figure}[b]
    \centering
    \subfloat[ASTRA\label{fig:mebt-astra}]{
        \includegraphics[width=0.48\linewidth]{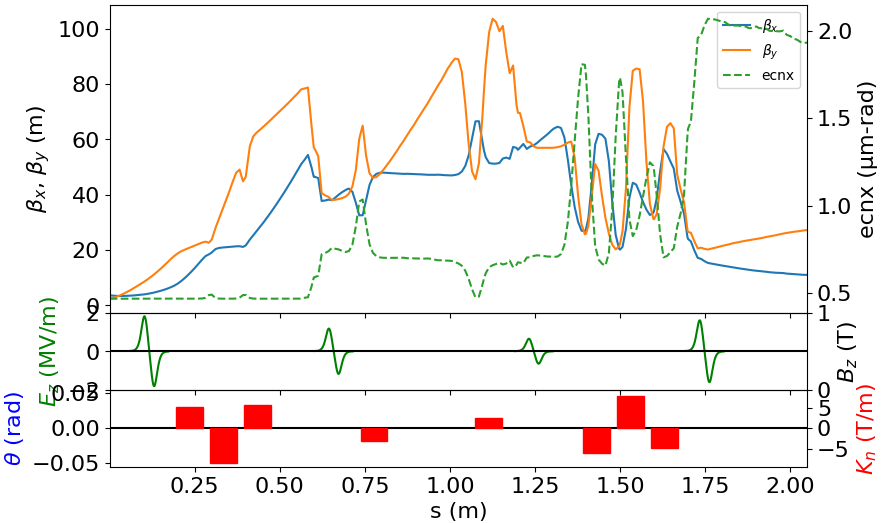}
    }
    \hfill
    \subfloat[Machine learning model\label{fig:mebt-poly}]{
        \includegraphics[width=0.48\linewidth]{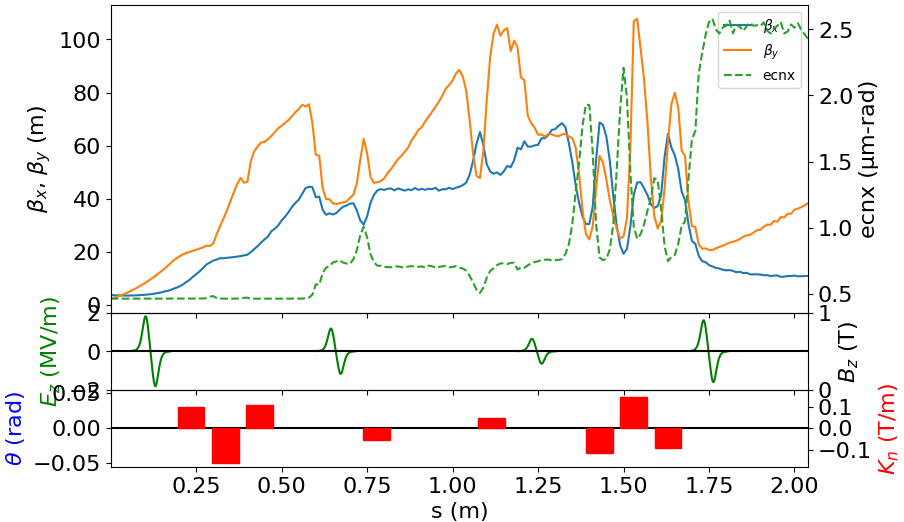}
    }
    \caption{Results of tracking the ISIS MEBT section through the DT using two different methods: a space-charge tracking code (ASTRA) and a machine learning model called via \texttt{Poly-Lithic}. In both cases, the machine was set and read entirely via the virtual control system, including the tracking method used.}
    \label{fig:mebt}
\end{figure}

The tracking code for a given machine section is available as a PV in the virtual control system, and if this is changed, the value is then sent to \texttt{SIMBA} for tracking. As mentioned above, \texttt{SIMBA} treats the machine learning model as any other simulation code; it can query the required inputs, check that they correspond to the current machine section, and structure the outputs accordingly. These outputs are then used to update the virtual control system, as is done with other tracking codes. The results taken from the virtual control system after tracking through the MEBT using ASTRA and the machine learning model, with the same machine settings and input parameters, are shown in Fig.\,\ref{fig:mebt}. Perhaps unsurprisingly, the agreement between the two sets of results are not perfect: the simple neural network was trained only for the purposes of demonstrating the functionality of the DT for switching seamlessly between different tracking methods. As more advanced machine learning models are deployed, these can be included as callable methods for the DT. The ISIS control system is currently undergoing a transition from Vsystem to EPICS \cite{IPAC2022.TUPOPT063}, so direct comparisons between the model and experimental measurements are not yet available; if the user would prefer to use a machine learning model trained on real measured data instead of simulation outputs, \texttt{SIMBA} can be bypassed, and instead the DT could be used for predictive measurements and offline optimization of the machine based only on the control system data.

%\iffalse
\subsection{Virtual Commissioning of UK XFEL}\label{subsec:ukxfel}
To demonstrate the utility of our DT architecture for offline testing of procedures and optimization across a range of project stages, a digital twin has been generated for the proposed UK XFEL facility and a virtual commissioning experiment has been developed. The injector \cite{PhysRevAccelBeams.28.091602}, main linac and bunch compressors \cite{arXiv.2603.08318}, and one of the free-electron laser (FEL) lines are simulated using the OPAL \cite{arXiv.1905.06654}, ELEGANT \cite{APS-LS-287} and GENESIS \cite{NIMA.429.1-3.243} codes, respectively, with the FEL simulation making use of a reduced model (the `steady-state' mode) in order to speed up calculations. A single line of the multi-FEL beam distribution switchyard is considered for simplicity.
Given that the facility is still in the design stage, control system variables for magnets and RF cavities were procedurally generated for the purposes of virtual commissioning, which could form the basis of future variable allocations.

\begin{figure}[h]
    \centering
    \fbox{\includegraphics[width=0.85\linewidth]{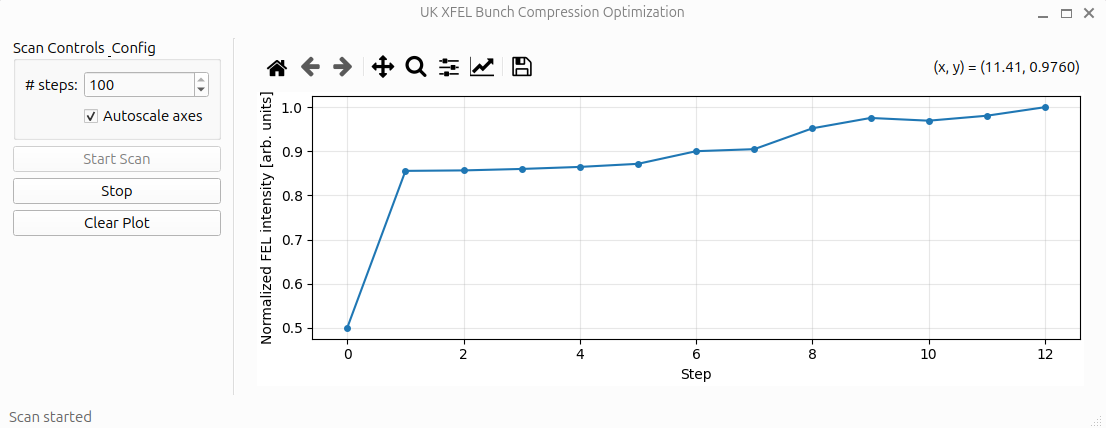}}
    \caption{Virtual optimization of the FEL intensity from the UK XFEL. Magnetic strengths were varied (via virtual PVs) to optimize FEL emission, with the data read from GENESIS output files and sent to the virtual control system.}
    \label{fig:ukxfel-gui}
\end{figure}

Figure\,\ref{fig:ukxfel-gui} shows a simple application optimizing the FEL intensity via optics matching and undulator settings. As with the other examples, this optimization was performed entirely in the virtual control system which was built automatically based on the PVs defined in the lattice file. Similarly, the simulation models for the various machine sections were also constructed from the elements defined in the same place. 
Lattice parameters were scanned to vary bunch compression and the FEL intensity was read from the simulation output files and written to the virtual control system at the end of each tracking loop. While some additional control information would need to be added to this optimization before porting this tool directly onto a physical accelerator, this procedure was performed entirely via control system interactions, demonstrating that this DT can easily be used for prototyping software applications and experimental procedures, even ahead of facility construction.

%\fi
\section{Conclusions}\label{sec:conclusions}

This article has described the structure and initial deployment of a generalizable digital twin for particle accelerator monitoring and control. The backbone of this architecture is provided by a generic description of the accelerator lattice that incorporates the information required for both constructing and interacting with a virtual accelerator, and building, executing and analyzing simulations. Using this as the ground source of truth about the lattice not only offers coherence to the entire structure of the DT, but also enables seamless integration with the physical control system. The modularity of the system is also beneficial: while the specific implementation described here is useful as a proof-of-concept, and has demonstrated promising results for the CLARA accelerator, the ISIS virtual injector and UK XFEL, other users may prefer to swap out certain modules depending on their needs. In principle, the containerized nature of this DT offers this flexibility. 

Two important aspects are required for further development in order to push this towards an advanced, fully integrated DT:
\begin{itemize}
    \item Depending on the user configuration, the update loop takes some time to execute, typically longer than the machine repetition rate. This is in part a result of the simulation method used. For our initial testing purposes, only a small number of particles ($4096$) are used, in order to reduce the computational overhead of passing full beam distributions between containers, and to speed up particle tracking simulations. For a relatively small machine such as CLARA or the ISIS injector, this reduces the execution time to a few seconds, provided that the low-energy injector is already simulated; larger facilities may experience longer latency. Further speed-up can be achieved in a variety of ways: 
    \begin{itemize}
        \item Employing reliable ML models via \texttt{Poly-Lithic} where possible.
        \item Running particle tracking simulations on a cluster -- an option that can be set in \texttt{SIMBA} -- or deploying the entire DT on a cluster.
        \item Reducing the number of particles in the beam distributions that are sent from the simulation module to the communications layer -- or indeed, dispensing with the full beam distributions entirely.
    \end{itemize}
    \item The results from the simulation must be carefully cross-checked with measured beam parameters before bi-directional \textit{control} between the physical accelerator and the DT is permitted. There is no generic solution to this, and each facility must first ensure that the results are reliable before allowing the DT to adjust the accelerator hardware. Typically, particle tracking codes produce an idealized representation of the physical system, while ML models can require careful tuning and training, and are prone to errors if they encounter a set of parameters that are outside of their training set.  
\end{itemize}

Some of these issues must be tackled on a case-by-case basis, however this architecture provides a significant step in the realization of DT technology that can be used across the accelerator community. While some preparatory work is necessary to create a lattice representation in \texttt{LAURA} format with the appropriate control system information, once this is done almost the entire DT can be procedurally generated and developed further for a specific use-case. The ability to swap out modules depending on the user's goals means that each facility-specific implementation can leverage their own methods, knowledge and experience to optimize this tool for different purposes. For example, the simulation engine could be excluded, and instead a model trained on control system data could be called to update the virtual accelerator. Alternatively, the virtual control system may be bypassed if the user's goal is simply to generate a database of indexed and searchable simulation information in order to train a machine learning model. 

This article has demonstrated the proof-of-principle of an auto-generating, configurable and extendable DT architecture that can be applied to many accelerator facilities. %The sharing of solutions to common challenges across the community is necessary in order to accelerate this effort, and to develop a generic DT solution for accelerators that can provide widespread benefits in terms of optimized control, predictive maintenance and operational flexibility. 
There is now an opportunity for a co-ordinated community effort to accelerate this work, through exchanging solutions to shared challenges, to achieve an advanced, fully integrated DT for accelerators that can provide widespread benefits in terms of optimized control, predictive maintenance and operational flexibility.

\bibliographystyle{unsrtnat}
\bibliography{references}

\end{document}